\documentclass{amsart}

\newtheorem{thm}{Theorem}[section]

\newtheorem{lem}[thm]{Lemma}

\theoremstyle{definition}

\theoremstyle{remark}

\numberwithin{equation}{section}

\newcommand{\Real}{\mathbb R}

\newcommand{\To}{\longrightarrow}

\begin{document}

\title[Bound states of two-dimensional Schr\"{o}dinger-Newton equations]{Bound states of two-dimensional Schr\"{o}dinger-Newton equations}%
\author{Joachim Stubbe}%
\address{EPFL, IMB-FSB, Station 8, CH-1015 Lausanne, Switzerland}
\email{Joachim.Stubbe@epfl.ch}%

\thanks{I am are indebted to Marc Vuffray for providing us his numerical studies of the model.
I also wish to thank Philippe Choquard for many helpful discussins
during the preparation of this paper.}

\subjclass{35Q55, 35Q40,47J10} \keywords{Schr\"{o}dinger-Newton
equations, nonlinear Schr\"{o}dinger equation, rearrangement
inequality, logarithmic Hardy-Littelwood-Sobolev inequality}

\date{06th February 2008}%
\begin{abstract}
We prove an existence and uniqueness result for ground states and
for purely angular excitations of two-dimensional
Schr\"{o}dinger-Newton equations. From the minimization problem
for ground states we obtain a sharp version of a logarithmic
Hardy-Littlewood-Sobolev type inequality.
\end{abstract}
\maketitle
\section{Introduction}
We consider the Schr\"{o}dinger-Newton system
\begin{equation}\label{SN}
    iu_t+\Delta u-\gamma Vu=0,\quad \Delta V=|u|^2
\end{equation}
in two space dimensions which is equivalent to the nonlinear
Schr\"{o}dinger equation
\begin{equation}\label{NLS}
    iu_t+\Delta u-\frac{\gamma}{2\pi}(\ln(|x|)*|u|^2)u=0
\end{equation}
with nonlocal nonlinear potential
\begin{equation*}
   V(x)=\frac1{2\pi} (\ln |x|*|u|^2)(x,t)=\frac1{2\pi}\int_{\Real^2}\ln (|x-y|)\,|u(t,y)|^2\;dy.
\end{equation*}
We are interested in the existence of nonlinear bound states of
the form
\begin{equation}\label{Bound states}
    u(t,x)=\phi_{\omega}(x)e^{-i\omega t}.
\end{equation}
The Schr\"{o}dinger-Newton system \eqref{SN} in three space
dimensions has a long standing history. With $\gamma$ designating
appropriate positive coupling constants it appeared first in 1954,
then in 1976 and lastly in 1996 for describing the quantum
mechanics of a Polaron at rest by S. J. Pekar ~\cite{P54}, of an
electron trapped in its own hole by  Ph. Choquard ~\cite{L77} and
of selfgravitating matter by R. Penrose ~\cite{P96}. The
two-dimensional model is studied numerically in ~\cite{HMT2003}.
For the bound state problem there are rigorous results only for
the three dimensional model. In ~\cite{L77} the existence of a
unique ground state of the form \eqref{Bound states} is shown by
solving an appropriate minimization problem. This ground state
solution $\phi_{\omega}(x), \omega<0$ is a positive spherically
symmetric strictly decreasing function. In ~\cite{L80} the
existence of infinitely many distinct spherically symmetric
solutions is proven and, in ~\cite{L86}, a proof for the existence
of anisotropic bound states is claimed. So far, there are no
results for the two-dimensional model using variational methods.
One mathematical difficulty of the two-dimensional problem is that
the Coulomb potential in two space dimensions is neither bounded
from above nor from below and hence does not define a positive
definite quadratic form. However, recently in ~\cite{CSV2008} the
existence of a unique positive spherically symmetric stationary
solution $(u,V)$ of \eqref{SN} such that $V(0)=0$ has been proven
by applying a shooting method to the corresponding system of
ordinary differential equations.

In the present paper, we are mainly interested in the ground
states of the model
\begin{equation}\label{Ground states}
    u(t,x)=\phi_{\omega}(x)e^{-i\omega t},\quad
    \phi_{\omega}(x)>0.
\end{equation}
We prove the existence of ground states by solving an appropriate
minimization problem for the energy functional $E(u)$.  By use of
strict rearrangement inequalities we shall prove that there is a
unique minimizer (up to translations and a phase factor) which is
a positive spherically symmetric decreasing function (theorems
\ref{existence-gs} and theorem \ref{uniqueness-gs}). The existence
and uniqueness of solutions of the form \eqref{Ground states} for
given $\omega$ ,however, depends on the frequency $\omega$. It
will turn out that for any $\omega \leq 0$ there is a unique
ground state of the form \eqref{Ground states}. In addition, we
shall prove there is a positive number $\omega^*>0$ independent of
the coupling $\gamma$ such that for any positive $\omega<\omega^*$
there are two ground states of the form \eqref{Ground states} with
different $L^2$-norm. In the limit case $\omega=\omega^*$ there is
again a unique ground state (theorem \ref{frequency-thm}). We then
apply our existence result to obtain a sharp
Hardy-Littlewood-Sobolev inequality for the logarithmic kernel
(theorem \ref{log-HLS-ineq-thm}).

Finally, we prove the existence of non radial solutions of the
Schr\"{o}dinger-Newton system \eqref{SN} which in polar
coordinates $(r,\theta)$ are of the form
\begin{equation}\label{m-ground-states}
    \psi(t,r,\theta)=\phi_{m,\omega}(r)e^{im\theta-i\omega t},\quad
    \phi_{m,\omega}(r)\geq 0
\end{equation}
for positive integers $m$. These are eigenfunctions of the angular
momentum operator $L=-i\partial_{\theta}$. Again, we prove the
existence of such $\phi_{m,\omega}(r)$ by solving an appropriate
minimization problem for the corresponding energy functional
reduced to functions of the form \eqref{m-ground-states} for any
given $m$ (theorem \ref{existence-gs-m}). The minimizers can be
interpreted as purely angular excitations and we also prove their
uniqueness in this class of functions (theorem
\ref{uniqueness-gs-m}).

\section{Mathematical Framework}

\subsection{Functional Setting} The natural function space $X$ for the quasi-stationary problem is given by
\begin{equation}\label{X}
    X=\{u:\mathbb{R}^2\to \mathbb{C}:\;\int_{\Real^2}|\nabla u|^2+|u|^2+\ln(1+|x|)\,|u|^2\;dx<\infty\}.
\end{equation}
The space $X$ is a Hilbert space and by Rellich's criterion (see,
e.g. theorem XIII.65 of ~\cite{RS4}) the embedding
$X\hookrightarrow L^2$ is compact. We note $X_r$ the space of
radial functions in $X$. Formally, the energy $E$ associated to
\eqref{NLS} is given by
\begin{equation}\label{energy}
\begin{split}
    E(u)&=\int_{\Real^2}|\nabla u(x)|^2\;dx+\frac{\gamma}{4\pi}\int_{\Real^2}\int_{\Real^2}\ln(|x-y|)\,|u(x)|^2|u(y)|^2\;dxdy\\
    &=T(u)+\frac{\gamma}{2}V(u)\\
    \end{split}
\end{equation}
In order to prove that the energy is indeed well defined on $X$ we
decompose the potential energy $V(u)$ into two parts applying the
identity
\begin{equation}\label{ln}
    \ln (r) =-\ln(1+\frac1{r})+\ln(1+r)
\end{equation}
for all $r>0$. We then define the corresponding functionals
\begin{equation}\label{V1V2}
\begin{split}
    &V_1(u)=-\frac{1}{2\pi}\int_{\Real^2}\int_{\Real^2}\ln(1+\frac1{|x-y|})\,|u(x)|^2|u(y)|^2\;dxdy\\
    &V_2(u)=\frac{1}{2\pi}\int_{\Real^2}\int_{\Real^2}\ln(1+{|x-y|})\,|u(x)|^2|u(y)|^2\;dxdy\\
\end{split}
\end{equation}
\begin{lem}
The energy functional $E:X\To\Real_0^{+}$ is well defined on $X$
and of class $C^1$.
\end{lem}
\begin{proof}
Since $0\leq \ln(1+\frac1{r})\leq \frac1{r}$ for any $r>0$ we have
by the Hardy-Littlewood-Sobolev inequality (see e.g.~\cite{L83})
and Sobolev interpolation estimates (see e.g.~\cite{W83})that
\begin{equation*}
    |V_1(u)|\leq C_1 ||u||^4_{8/3}\leq C_2 ||\nabla
    u||_2||u||_2^3
\end{equation*}
for some constants $C_1,C_2>0$. To bound the second term of the
potential energy we note that
\begin{equation*}
    \ln(1+|x-y|)\leq \ln (1+|x|+|y|)\leq \ln (1+|x|)+ \ln (1+|y|)
\end{equation*}
and therefore
\begin{equation*}
    |V_2(u)|\leq \frac{1}{\pi}\int_{\Real^2}\ln(1+|x|)\,u(x)|^2\int_{\Real^2}|u(y)|^2\;dxdy.\\
\end{equation*}
The regularity properties of $E(u)$ are obvious.
\end{proof}
Finally, we consider the particle number (or charge) defined by
\begin{equation}\label{charge}
    N(u)=\int_{\Real^2}|u(x)|^2\;dx,
\end{equation}
which is also a well-defined quantity on $X$.

\subsection{Scaling properties} If $\phi_{\omega}(x)$ is a
solution of the stationary equation
\begin{equation}\label{sNLS-omega}
    -\Delta \phi_{\omega}(x)+\frac{\gamma}{2\pi}\bigg(\int_{\Real^2}\ln(|x-y|)\;|\phi_{\omega}(y)|^2\;dy\bigg)\;\phi_{\omega}(x)=\omega\phi_{\omega}(x),
\end{equation}
with finite particle number $N_{\omega}=N(\phi_{\omega})$ then by
the virial theorem,
\begin{equation}\label{virial}
    T(\phi_{\omega})=\frac{\gamma}{8\pi}N^2(\phi_{\omega}).
\end{equation}
For any $\sigma>0$, the scaled function
$\phi_{\omega,\sigma}(x)=\sigma^{-2}\gamma^{1/2}\phi_{\omega}({x}/{\sigma})$
solves
\begin{equation}\label{sNLS-1}
\begin{split}
    &-\Delta\phi_{\omega,\sigma}(x)+\frac{1}{2\pi}\bigg(\int_{\Real^2}\ln
    (|x-y|)\;|\phi_{\omega,\sigma}(y)|^2\;dy\bigg)\;\phi_{\omega,\sigma}(x)\\
    &=\\
    &\sigma^{-2}\big(\omega+\frac{\gamma
    N_{\omega}\ln\sigma}{2\pi}\big)\phi_{\omega,\sigma}(x).\\
    \end{split}
\end{equation}
If we choose $\sigma=\sigma_{\omega}$ such that
$\omega+\frac{\gamma
    N_{\omega}\ln\sigma_{\omega}}{2\pi}=0$, then
    $\phi:=\phi_{\omega,\sigma_{\omega}}$ is independent of
    $\omega, \gamma$ and satisfies the Schr\"{o}dinger equation
\begin{equation}\label{sNLS-universal}
-\Delta\phi(x)+\frac{1}{2\pi}\bigg(\int_{\Real^2}\ln
    (|x-y|)\;|\phi(y)|^2\;dy\bigg)\;\phi(x)=0.
\end{equation}
Define
\begin{equation}\label{LAMBDA-0}
    \Lambda_0:=N(\phi).
\end{equation}

Then the virial theorem \eqref{virial} reads
\begin{equation}\label{virial-0}
    T(\phi)=\frac{\Lambda_0^2}{8\pi}.
\end{equation}
From \eqref{sNLS-universal} we get $V(\phi)=-T(\phi)$, hence
\begin{equation}\label{scaling-universal}
    E(\phi)=T(\phi)+\frac1{2}V(\phi)=\frac{\Lambda_0^2}{16\pi}.
\end{equation}
The scaling of the particle number
\begin{equation*}
    \Lambda_0=\sigma_{\omega}^{-2}\gamma N_{\omega}
\end{equation*}
yields a relation between frequency $\omega$ and the particle
number $N_{\omega}$ which we shall discuss in detail in the
following section for ground state solutions.
\section{Ground states}
\subsection{Existence of ground states} We consider the following
minimization problem:
\begin{equation}\label{mini0}
    e_0(\lambda)=\inf \{E(u),u\in X,N(u)=\lambda\}.
\end{equation}
We note that the functional $u\to E(u)$ is not convex since the
quadratic form $f\to
\int_{\Real}\int_{\Real}|x-y|f(x)\bar{f}(y)\;dxdy$ is not positive
so that standard convex minimization does not apply. We shall
prove the following theorem:
\begin{thm}\label{existence-gs} For any $\lambda>0$ there is a spherically symmetric
decreasing $u_{\lambda}\in X$ such that
$e_0(\lambda)=E(u_{\lambda})$ and $N(u_{\lambda})=\lambda$.
\end{thm}
\begin{proof} Let $(u_n)_n$ be a minimizing sequence for
$e_0(\lambda)$, that is $N(u_n)=\lambda$ and
$\underset{n\To\infty}{\lim} E(u_n)=e_0(\lambda)$. We also may
assume that $E(u_n)$ is uniformly bounded above. Denoting $u^{*}$
the spherically symmetric-decreasing rearrangement of $u$ we have
(see e.g. lemma 7.17 in ~\cite{LL01})
\begin{equation*}
    T(u)\geq T(u^{*}), \quad N(u^{*})= N(u).
\end{equation*}
Applying the decomposition $V(u)=V_1(u)+V_2(u)$ defined in
\eqref{V1V2} we may apply the strict version of Riesz's
rearrangement inequality (see e.g. theorem 3.9 in ~\cite{LL01}) to
$V_1(u)$ since $\ln(1+1/|x|)$ is positive and strictly
symmetric-decreasing. Therefore
\begin{equation*}
    V_1(u)\geq V_1(u^{*})
\end{equation*}
with equality only if $u(x)=u^{*}(x-x_0)$ for some
$x_0\in\mathbb{R}^2$. For the second term $V_2(u)$ we apply the
following rearrangement inequality:
\begin{lem}
Let $f,g$ be two nonnegative functions on $\Real$, vanishing at
infinity with spherically symmetric-decreasing rearrangement
$f^*,g^*$ , respectively. Let $v$ be a nonnegative spherically
symmetric increasing function. Then

\begin{equation}\label{Riesz2}
    \int_{\Real^2}\int_{\Real^2}f(x)v(x-y)g(y)\;dxdy\geq \int_{\Real^2}\int_{\Real^2}f^*(x)v(x-y)g^*(y)\;dxdy
\end{equation}
\end{lem}
\begin{proof}
The proof follows the same lines as in ~\cite{CS2007}, lemma 3.2,
where we proved the corresponding lemma in one space dimension. We
give it here for the sake of completeness. If $v$ is bounded,
$v\leq C$, then $(C-v)^*=C-v$ and by Riesz's rearrangement
inequality (lemma 3.6 in ~\cite{LL01}) we have
\begin{equation*}
    \int_{\Real^2}\int_{\Real^2}f(x)(C-v(x-y))g(y)\;dxdy\leq
    \int_{\Real^2}\int_{\Real^2}f^{*}(x)(C-v(x-y))g^{*}(y)\;dxdy.
\end{equation*}
Since
\begin{equation*}
    \int_{\Real^2}f(x)\;dx\int_{\Real^2}g(y)\;dy =
    \int_{\Real^2}f^{*}(x)\;dx\int_{\Real^2}g^{*}(y)\;dy
\end{equation*}
the claim follows. If $v$ is unbounded we define a truncation by
$v_n(x)=\sup{(v(x),n)}$ and apply the monotone convergence
theorem.
\end{proof}
By the preceding lemma we have
\begin{equation*}
    V_2(u)\geq  V_2(u^{*})
\end{equation*}
and consequently
\begin{equation*}
    V(u)\geq  V(u^{*})
\end{equation*}
with equality only if $u(x)=e^{i\theta} u^{*}(x-x_0)$ for some
$\theta \in\mathbb{R}$ and $x_0\in\mathbb{R}^2$. Therefore we may
suppose that $u_n=u^{*}_n$. We claim that $u^{*}_n\in X$. Indeed,
by Newton's theorem (see e.g. theorem 9.7 in ~\cite{LL01}) we have
\begin{equation}\label{Newton-theorem}
\begin{split}
    V(u^{*}_n)&=\frac{1}{2\pi}\int_{\Real^2}{u^{*}_n(x)}^2(\ln|x|)\;\bigg(\int_{B_{|x|}}{u^{*}_n(y)}^2\;dy\bigg)\;dx\\
    &+\frac{1}{2\pi}\int_{\Real^2}{u^{*}_n(x)}^2\bigg(\int_{\Real^2\setminus
    B_{|x|}}(\ln|y|)\;{u^{*}_n(y)}^2\;dy\bigg)\;dx\\
    \end{split}
\end{equation}
where $B_{|x|}$ denotes the disc of radius $|x|$ centered at the
origin. Since  $\ln |y|\geq \ln|x|$ for all $y\in\Real^2\setminus
B_{|x|}$ we get
\begin{equation*}
     V(u^{*}_n)\geq
     \frac{\lambda}{2\pi}\int_{\Real^2}(\ln|x|)\;{u^{*}_n(x)}^2\;dx.
\end{equation*}
Using
\begin{equation*}
    \ln |x|\geq \ln(1+|x|)-\frac1{|x|}
\end{equation*}
and the sharp Sobolev inequality between the linear operators
$-\Delta$ and $\frac1{|x|}$
\begin{equation}\label{op-ineq}
   \int_{\Real^2}\frac1{|x|}\;|u(x)|^2\;dx\leq 2||u||_2||\nabla
   u||_2
\end{equation}
we finally get
\begin{equation*}
     V(u^{*}_n)\geq
     \frac{\lambda}{2\pi}\int_{\Real^2}\ln(1+|x|)\;{u^{*}_n(x)}^2\;dx-\frac{\lambda^{3/2}}{\pi}||\nabla
   u^{*}_n||_2.
\end{equation*}
Hence
\begin{equation*}
    E(u^{*}_n)\geq
    ||\nabla u^{*}_n||_2^2+\frac{\gamma\lambda}{4\pi}\int_{\Real^2}\ln(1+|x|)\;{u^{*}_n(x)}^2\;dx-\frac{\gamma\lambda^{3/2}}{2\pi}||\nabla
   u^{*}_n||_2
\end{equation*}
proving our claim. We may extract a subsequence which we denote
again by $(u^{*}_n)_n$ such that $u^{*}_n\to u^{*}$ weakly in $X$,
strongly in $L^2$ and a.e. where $u^{*}\in X$ is a nonnegative
spherically symmetric decreasing function. Note that $u^{*}\neq 0$
since $N(u^{*})=\lambda$. We want to show that $E(u^{*})\leq
\underset{n\To\infty}{\lim\inf}\;E(u^{*}_n)$. Since
\begin{equation*}
    T(u^{*})\leq
\underset{n\To\infty}{\lim\inf}\;T(u^{*}_n)
\end{equation*}
it remains to analyze the functional $V(u)$. Let
\begin{equation*}
    \eta(x)=\int_{B_{|x|}}|u^{*}(y)|^2\;dy,\quad
    \eta_n(x)=\int_{B_{|x|}}|u^{*}_{n}(y)|^2\;dy.
\end{equation*}
Then $\eta_n(x)\to\eta(x)$ uniformly since
\begin{equation*}
    ||\eta_n(x)-\eta(x)||_{\infty}\leq  ||u^{*}_n-u^{*}||_{2}\;
    ||u^{*}_n+u^{*}||_{2}\leq
    2\sqrt{\lambda}||u^{*}_n-u^{*}||_{2}.
\end{equation*}
We note that for any spherically symmetric density $|u(x)|^2$ with
$u\in X$ we may simplify \eqref{Newton-theorem} to
\begin{equation*}
    V(u^{*}_n)=\frac{1}{\pi}\int_{\Real^2}{u^{*}_n(x)}^2(\ln|x|)\;\bigg(\int_{B_{|x|}}{u^{*}_n(y)}^2\;dy\bigg)\;dx.
\end{equation*}
Therefore, by the definition of $\eta_n,\eta$, we have
\begin{equation}\label{V-wlsc}
\begin{split}
   V(u^{*}_n)-V(u^{*})&=\frac{1}{\pi}\int_{\Real^2}(\ln|x|)|u^{*}_n(x)|^2\big(\eta_n(x)-\eta(x)\big)\;dx\\
   &\quad +\frac{1}{\pi}\int_{\Real^2}(\ln|x|)\eta(x)\big(|u^{*}_n(x)|^2-|u^{*}(x)|^2\big)\;dx\\
   \end{split}
\end{equation}
As $n\to\infty$ the first integral in \eqref{V-wlsc} will tend to
zero. In order to analyze the second integral we again decompose
$\ln |x|$ according to \eqref{ln}. Then
\begin{equation*}
    \frac{1}{\pi}\int_{\Real^2}\ln(1+|x|)\eta(x)\big(|u^{*}_n(x)|^2-|u^{*}(x)|^2\big)\;dx
\end{equation*}
will remain nonnegative since the continuous functional
$\phi\to\int_{\Real^2}\ln(1+|x|)\eta(x)|\phi(x)|^2\;dx$ is
positive while
\begin{equation*}
    -\frac{1}{\pi}\int_{\Real^2}\ln(1+\frac1{|x|})\eta(x)\big(|u^{*}_n(x)|^2-|u^{*}(x)|^2\big)\;dx
\end{equation*}
converges to zero since by H\"{o}lder's inequality and inequality
\eqref{op-ineq} we have the estimate
\begin{equation*}
\begin{split}
    &\frac{1}{\pi}\int_{\Real^2}\ln(1+\frac1{|x|})\eta(x)\big(|u^{*}_n(x)|^2-|u^{*}(x)|^2\big)\;dx\\
    &\leq \frac{2}{\pi} ||\eta||_{\infty}||\nabla (u^{*}_n+u^{*})||_2^{1/2}||u^{*}_n+u^{*}||_2^{1/2}||\nabla
    (u^{*}_n-u^{*})||_2^{1/2}||u^{*}_n-u^{*}||_2^{1/2}\\
    &\leq C ||u^{*}_n-u^{*}||_2^{1/2}\\
\end{split}
\end{equation*}
for a positive constant $C$. Hence
\begin{equation*}
    V(u^{*})\leq
\underset{n\To\infty}{\lim\inf}\;V(u^{*}_n)
\end{equation*}
proving the theorem.
\end{proof}
\subsection{Uniqueness of ground states} Let $u_{\lambda}$ denote
the solution of the minimization problem \eqref{mini0} found in
theorem \label{existence-gs}.
\begin{thm}\label{uniqueness-gs}
For any $\lambda>0$ the solution $u_{\lambda}$ of the minimization
problem \eqref{mini0} is unique in the following sense: If
$v_{\lambda}\in X$ is such that $e_0(\lambda)=E(v_{\lambda})$ and
$N(v_{\lambda})=\lambda$, then $v_{\lambda}\in\{ e^{i\theta}
u^{*}(x-x_0),\theta \in\mathbb{R},x_0\in\mathbb{R}^2\}$.
\end{thm}
\begin{proof}
Since by the strict rearrangement inequality $E(u)>E(u^{*})$ for
all $u\notin\{ e^{i\theta} u^{*}(x-x_0),\theta
\in\mathbb{R},x_0\in\mathbb{R}^2\}$ established in the proof of
theorem \ref{existence-gs} it is sufficient to prove uniqueness in
the class of spherically symmetric (decreasing) nonnegative
functions. To do so we consider the Euler-Lagrange equation for
$u_{\lambda}$. By theorem \ref{existence-gs} for any $\lambda>0$
there is a Lagrange multiplier $\omega$ such that, at least in a
weak sense
\begin{equation}\label{Euler-Lagrange-eq}
    -\Delta u_{\lambda}(x)+\frac{\gamma}{2\pi}\bigg(\int_{\Real^2}\ln(|x-y|)\;|u_{\lambda}(y)|^2\;dy\bigg)\;u_{\lambda}(x)=\omega
    u_{\lambda}(x).
\end{equation}
We define the quantities
\begin{equation*}
    I_{\lambda}:=\frac{1}{2\pi}\int_{\Real^2}\ln(|y|)\;|u_{\lambda}(y)|^2\;dy,\quad
    E_{\lambda}:=\omega-\gamma I_{\lambda}
\end{equation*}
and the potential
\begin{equation*}
    W_{\lambda}(x):=\frac{1}{2\pi}\bigg(\int_{\Real^2}\ln(|x-y|)\;|u_{\lambda}(y)|^2\;dy\bigg)-I_{\lambda}.
\end{equation*}
Then $W_{\lambda}(x)\geq 0$ for all $x\in\mathbb{R}^2$,
$W_{\lambda}(0)=0$ and $\Delta W_{\lambda}=|u_{\lambda}|^2$. The
function $u_{\lambda}$ is the ground state of the Schr\"{o}dinger
operator $-\Delta+\gamma W_{\lambda}$ with eigenvalue
$E_{\lambda}$ and therefore $E_{\lambda}>0$. The rescaled
functions
\begin{equation}\label{scaling-universal-system}
u(|x|):=\frac{\sqrt{\gamma}}{E_{\lambda}}\;u_{\lambda}(x/\sqrt{E_{\lambda}}),\quad
W(|x|):=\frac{{\gamma}}{E_{\lambda}}\;W_{\lambda}(x/\sqrt{E_{\lambda}})
\end{equation}
then $u,W$ satisfy the universal equations
\begin{equation}\label{universal-system}
    \Delta u=(W-1)u,\quad \Delta W=|u|^2.
\end{equation}
In ~\cite{CSV2008} it was shown that \eqref{universal-system}
admits a unique spherically symmetric solution $(u,W)$ such that
$u$ is a positive decreasing function vanishing at infinity. See
also theorem \ref{uniqueness-gs-m} of the present paper where we
consider a more general system.
\end{proof}
\subsection{Dependence on parameters and admissible frequencies} We determine explicitly $e_0(\lambda)$ as a function of $\lambda$ and the range of
admissible frequencies $\omega$ for ground states of the form
\eqref{Ground states}. We consider the unique spherically
symmetric solution $(u,W)$, $u>0$ and $u$ vanishing at infinity ,
of the universal system \eqref{universal-system}. Let
\begin{equation}\label{N-I}
    N:=\int_{\Real^2}|u(|x|)|^2\;dx,\quad
    I:=\frac{1}{2\pi}\int_{\Real^2}\ln(|x|)\;|u(|x|)|^2\;dx.
\end{equation}
These quantities are finite since, for any $p>0$, $u$ decays
faster than $\exp(-px)$ at infinity and in particular $u\in X$.
The corresponding numerical values are easily computed from the
solution of \label{universal-system} and we give them in the
appendix. By the scaling \eqref{scaling-universal-system} we
obtain the following identity for the parameters $\gamma, \lambda$
and the Lagrange multiplier $\omega$:
\begin{equation}\label{gamma-omega-lambda-relation}
    \omega=\frac{\gamma\lambda}{N}\big(1+I-\frac{N}{4\pi}\ln\frac{\gamma\lambda}{N}\big).
\end{equation}
Multiplying the variational equation \eqref{Euler-Lagrange-eq} by
$u_{\lambda}$ and integrating we obtain
\begin{equation*}
    2E(u_{\lambda})-T(u_{\lambda})=T(u_{\lambda})+\gamma V(u_{\lambda})=\omega\lambda.
\end{equation*}
Applying the virial relation \eqref{virial} and using
\eqref{gamma-omega-lambda-relation} we finally have
\begin{equation}\label{explicit-e0}
    e_0(\lambda)=\frac{\gamma\lambda^2}{16\pi}\Big(1+8\pi\frac{1+I}{N}-2\ln\big(\frac{\gamma\lambda}{N}\big)\Big)
\end{equation}
with $N,I$ given in \eqref{N-I}. Note that \eqref{explicit-e0} can
be also obtained by integrating $\omega=\omega(\lambda)$ in
\eqref{gamma-omega-lambda-relation} with respect to $\lambda$
since, at least formally,
\begin{equation*}
    \frac{d\,e_0(\lambda)}{d\lambda}=\omega.
\end{equation*}
Taking $\phi$ in \eqref{sNLS-universal} as the unique ground state
solution the relation \eqref{gamma-omega-lambda-relation} between
$\lambda,\omega$ and $\gamma$ simplifies to
\begin{equation}\label{gamma-omega-lambda-relation-Lambda0}
    \omega=-\frac{\gamma\lambda}{4\pi}\ln\frac{\gamma\lambda}{\Lambda_0}
\end{equation}
with $\Lambda_0=N(\phi)$ as defined in \eqref{LAMBDA-0}. The
energy is then given by
\begin{equation}\label{explicit-e0-effective}
    e_0(\lambda)=\frac{\gamma\lambda^2}{16\pi}\big(1-2\ln \frac{\gamma\lambda}{\Lambda_0}\big).
\end{equation}
The relations \eqref{gamma-omega-lambda-relation}, respectively
\eqref{gamma-omega-lambda-relation-Lambda0}, yield the following
result on admissible frequencies:
\begin{thm}\label{frequency-thm}

(1)  For any $\omega\leq 0$ the Schr\"{o}dinger-Newton system
\eqref{SN} admits a unique spherically ground state of the form
\eqref{Ground states}. In particular, for the solution $\phi$ of
the stationary equation \eqref{sNLS-universal} the energy
$e_0(\lambda)$ attains its maximum.

(2) For any $0<\omega < \omega^*$ with
\begin{equation*}
    \omega^*=\frac{N}{4\pi e}\exp\big(4\pi\;\frac{1+I}{N}\big)=\frac{\Lambda_0}{4\pi e}
\end{equation*}
the Schr\"{o}dinger-Newton system \eqref{SN} admits two
spherically ground states of the form \eqref{Ground states} with
different particle numbers.

(3) For $\omega=\omega^*$ there is a unique ground state solution
of the form \eqref{Ground states}.
\end{thm}

\section{A logarithmic Hardy-Littlewood-Sobolev inequality}
In ~\cite{CL92}, E. Carlen and M. Loss proved the following
logarithmic Hardy-Littlewood-Sobolev inequality:
\begin{equation}\label{log-HLS-CL92}
   - \int_{\Real^n}\int_{\Real^n}\ln(|x-y|)\,f(x)f(y)\;dxdy\leq
    \frac{1}{n}\int_{\Real^n}f(x)\ln f(x)\;dx+C_0
\end{equation}
for all nonnegative real-valued $f$ with
\begin{equation*}
    \int_{\Real^n}f(x)\;dx=1,\int_{\Real^n}f(x)\ln(1+|x|)\;dx,
\int_{\Real^n}f(x)\ln f(x)\;dx<\infty
\end{equation*}
and sharp constant $C_0$. We focus on the case $n=2$. Replacing
$f$ by $|u|^2$ with $N(u)=\lambda$ this reads as
\begin{equation}\label{log-HLS-CL92-2}
    -2\pi V(u)\leq
    \frac{\lambda}{2}\int_{\Real^n}|u(x)|^2\ln
    |u(x)|^2\;dx+C_0\lambda^2-\frac{\lambda^2}{2}\ln\lambda.
\end{equation}
Applying the logarithmic Sobolev inequality (~\cite{W78}, see also
~\cite{LL01}, theorem 8.14)
\begin{equation}\label{log-S}
    \int_{\Real^2}|u(x)|^2\ln |u(x)|^2\;dx\leq
    \lambda\ln(T(u))-\lambda(1+\ln\pi)
\end{equation}
to the integral on the r.h.s. of \eqref{log-HLS-CL92-2} and using
$2C_0=1+\ln\pi$ when $n=2$ we get
\begin{equation*}
    -V(u)\leq\frac{\lambda^2}{4\pi}\ln\frac{T(u)}{\lambda}.
\end{equation*}
However, this inequality is not sharp. Here we give the sharp
version:
\begin{thm}\label{log-HLS-ineq-thm} For any $u\in X$ the following inequality holds:
\begin{equation}\label{log-HLS-S08}
     -V(u)\leq\frac{\lambda^2}{4\pi}\ln\frac{8\pi T(u)}{N\lambda}-\big(\frac{I+1}{N}-\frac1{8\pi}\big)\lambda^2
\end{equation}
where $\lambda=\int_{\Real^2} |u|^2\;dx$ and the constants $I$ and
$N$ are given in \eqref{N-I}.
\end{thm}

\begin{proof} By theorem \ref{existence-gs} we have for any $u\in
X$ with $\lambda=\int_{\Real^2} |u|^2\;dx$ and for any coupling
constant $\gamma>0$ the sharp inequality
\begin{equation*}
    T(u)+\frac{\gamma}{2}V(u)\geq e_0(\lambda)
\end{equation*}
with $e_0(\lambda)$ given in \eqref{explicit-e0}. We optimize with
respect to $\gamma$. The optimal $\gamma$ is given by

\begin{equation*}
    \frac1{4\pi}\ln\frac{\gamma\lambda}{N}=-\lambda^{-2}V(u)+\frac{I+1}{N}-\frac1{8\pi}
\end{equation*}
which yields the desired inequality.
\end{proof}

\section{Purely angular excitations}
In this section we prove the existence of non radial solutions of
the Schr\"{o}dinger-Newton system \eqref{SN} which in polar
coordinates $(r,\theta)$ are of the form
\begin{equation*}
    \psi(t,r,\theta)=\phi_{\omega}(r)e^{im\theta-i\omega t},\quad
    \phi_{\omega}(r)\geq 0
\end{equation*}
for positive integers $m$. These are eigenfunctions of the angular
momentum operator $L=-i\partial_{\theta}$. We define the function
space $X_m^{(s)}$ of spherically symmetric functions
$u:\mathbb{R}^2\to \mathbb{C}$ by

\begin{equation}\label{X-m-radial}
    X_m^{(s)}=\{u=u(|x|): \int_{\Real^2}|\partial_r u|^2+\frac{m^2}{r^2}||u|^2+|u|^2+\ln(1+r)\,|u|^2\;dx<\infty\}.
\end{equation}
We define the energy functional for purely angular excitations by
\begin{equation}\label{energy-m}
    E_m(u)=T_m(u)+\frac{\gamma}{2}V(u),\quad T_m(u)=\int_{\Real^2}|\partial_r u|^2+\frac{m^2}{r^2}||u|^2\;dx,
\end{equation}
We consider the minimization problem
\begin{equation}\label{mini0}
    e_m(\lambda)=\inf \{E_m(u),u\in X_m^{(s)},N(u)=\lambda\}.
\end{equation}
The following theorem holds:
\begin{thm}\label{existence-gs-m} For any $\lambda>0$ there is a nonnegative $u_{\lambda}\in X_m^{(s)}$ such that
$e_m(\lambda)=E_m(u_{\lambda})$ and $N(u_{\lambda})=\lambda$.
\end{thm}
\begin{proof}
Consider any minimizing sequence $(u_n)_n$ for $e_m(\lambda)$.
Since $E_m(u)$ is decreasing under the replacement $u\mapsto |u|$
we may suppose that $u_n\geq 0$. Now we mimic the proof of theorem
\ref{existence-gs} starting from Newton's theorem
\eqref{Newton-theorem} replacing the energy $E$ by $E_m$ since
only the spherical symmetry of the functions are required in this
part of the proof.
\end{proof}
The minimizing $u_{\lambda}$ satisfies the Euler-Lagrange equation
\begin{equation}\label{Euler-Lagrange-eq}
    -u_{\lambda}''-\frac1{r}u'_{\lambda}+\frac{m^2}{r^2}u_{\lambda}+\frac{\gamma}{2\pi}\bigg(\int_{\Real^2}\ln(|x-y|)\;|u_{\lambda}(|y|)|^2\;dy\bigg)\;u_{\lambda}=\omega
    u_{\lambda}.
\end{equation}
for $r\geq 0$ and some multiplier $\omega$. By the same rescaling
as in the proof of theorem \ref{uniqueness-gs} we obtain a system
of universal equations given by
\begin{equation}\label{universal-system-m}
     u''+\frac1{r}u'-\frac{m^2}{r^2}u=(W-1)u,\quad W''+\frac1{r}W'=|u|^2
\end{equation}
such that $W(0)=W'(0)=0$ and $u\geq 0$. We note that
$f(r):=r^{-m}u(r)$ then satisfies $f(0)>0$, $f'(0)=0$ and the
differential equation
\begin{equation*}
    f''+\frac{2m+1}{r}f'=(W-1)f
\end{equation*}
from which we easily deduce $f>0$ and $f'<0$ using the facts
$f\geq 0$ and $f\to 0$ as $r\to\infty$. Also note that for any
$p>0$, $u$ decays faster than $\exp(-pr)$ at infinity. Uniqueness
of the solution $u_{\lambda}\in X_m^{(s)}$ follows again from the
uniqueness of the solutions of the universal system
\eqref{universal-system-m} which we prove in the following
theorem:
\begin{thm}\label{uniqueness-gs-m} For any $m\geq0$ there is a unique solution $(u, W)$ of
\eqref{universal-system-m} such that $u>0$ for $r>0$ and $u\to 0$
as $r\to\infty$.
\end{thm}
\begin{proof} Suppose there are two distinct solutions $(u_1,W_1)$,
$(u_2,W_2)$ having the required properties. We may suppose
$u_2(r)>u_1(r)$ for $r\in ]0,\bar{r}[$. We consider the Wronskian
\begin{equation*}
    w(r)=u_2'(r)u_1(r)-u_1'(r)u_2(r).
\end{equation*}
Note that $w(0)=0$ and $rw(r)\to 0$ as $r\to\infty$. It satisfies
the differential equation
\begin{equation*}
    (rw)'=r(W_2-W_1)u_1u_2.
\end{equation*}
Suppose $u_2(r)>u_1(r)$ for all $r>0$. Then $W_2(r)>W_1(r)$ for
all $r> 0$ since $r(W_2-W_1)'=\int_0^r (u_2^2-u_1^2)s\;ds>0$ and
hence $(rw)'>0$ for all $r> 0$ which is impossible. Hence there
exists $\bar{r}>0$ such that $\delta(r)=u_2(r)-u_1(r)>0$ for
$r\in[0,\bar{r}[$, $\delta(\bar{r})=0$ and $\delta'(\bar{r})<0$.
However, then $w(\bar{r})=\delta'(\bar{r})u_1(\bar{r})<0$, but
$(rw)'(r)>0$ for all $r< \bar{r}$, that is $w(r)>0$ for all $r<
\bar{r}$ which is again impossible.
\end{proof}

\appendix
\section{Numerical values}
Let $(u,W)$ be the unique spherically symmetric solution of the
universal equations \eqref{universal-system} such that
$W(0)=W'(0)=0$, $u>0$ vanishing at infinity, i.e.
\begin{equation}\label{universal-system-radialversion}
    (ru')'=r(W-1)u,\quad  (rW')'=r|u|^2.
\end{equation}
By integrating the second equation we get

\begin{equation}\label{N-eq}
    N=2\pi\int_0^{\infty}u^2(s)s\;ds=2\pi\underset{r\To\infty}{\lim}r
    W'(r).
\end{equation}
Multiplying the second equation by $\ln r$ and using
\begin{equation*}
    (rW')'\ln r=(rW'\ln r -W)'
\end{equation*}
we obtain after integration
\begin{equation}\label{I-eq}
    I=\int_0^{\infty}u^2(s)s\ln s\;ds=\underset{r\To\infty}{\lim}rW'(r)\ln r-W(r).
\end{equation}
Solving \eqref{universal-system-radialversion} numerically we then
find
\begin{equation}\label{N-I-numerical}
    N=2\pi\cdot 1.64145=10.3135,\quad I=0.2276
\end{equation}
and therefore
\begin{equation}\label{LAMBDA-0-numerical}
    \Lambda_0=46.03.
\end{equation}
With these numerical values the logarithmic
Hardy-Littlewood-Sobolev inequality \eqref{log-HLS-S08} reads as
follows:
\begin{equation}
     -V(u)\leq\frac{\lambda^2}{4\pi}\ln\frac{T(u)}{\lambda}-0.0084\lambda^2.
\end{equation}

\bibliographystyle{amsplain}

\end{document}